\documentclass[preprint2]{emulateapj}

\newcommand{\bzcat}{ROMA-BZCAT}

\newcommand{\fer}{{\it Fermi}}

\newcommand{\wse}{{\it WISE}}

\usepackage{graphicx}
\usepackage{longtable}
\usepackage{mdwlist}

\slugcomment{version \today: fm}
\shorttitle{Optical spectroscopic observations of blazars}
\shortauthors{F. Massaro et al. 2014}

\begin{document}
\title{Optical spectroscopic observations of blazars 
and $\gamma$-ray blazar candidates \\ in the Sloan Digital Sky Survey Data Release Nine}
\author{
F. Massaro\altaffilmark{1},
N. Masetti\altaffilmark{3},
R. D'Abrusco\altaffilmark{4},
A. Paggi\altaffilmark{4},
\&
S. Funk\altaffilmark{2}
}

\altaffiltext{1}{Yale Center for Astronomy and Astrophysics, Physics Department, Yale University, PO Box 208120, New Haven, CT 06520-8120, USA}
\altaffiltext{2}{SLAC National Laboratory and Kavli Institute for Particle Astrophysics and Cosmology, 2575 Sand Hill Road, Menlo Park, CA 94025, USA}
\altaffiltext{3}{INAF - Istituto di Astrofisica Spaziale e Fisica Cosmica di Bologna, via Gobetti 101, 40129, Bologna, Italy}
\altaffiltext{4}{Harvard - Smithsonian Astrophysical Observatory, 60 Garden Street, Cambridge, MA 02138, USA}

\begin{abstract}
We present an analysis of the optical spectra available in the Sloan Digital Sky survey data release nine (SDSS DR9) for the 
blazars listed in the \bzcat\ and for the $\gamma$-ray blazar candidates selected according to their IR colors.
First, we adopt a statistical approach based on MonteCarlo simulations 
to find the optical counterparts of the blazarslisted in the \bzcat\ catalog.
Then we crossmatched the SDSS spectroscopic catalog with our selected samples of blazars and $\gamma$-ray blazar candidates
searching for those with optical spectra available to classify our blazar-like sources and, whenever possible, to confirm their redshifts. 
Our main objectives are determining the classification of uncertain blazars listed in the \bzcat\ and discovering new gamma-ray blazars.
For the \bzcat\ sources we investigated a sample of 84 blazars confirming the classification for 20 of them and obtaining 18 new redshift 
estimates. For the $\gamma$-ray blazars, indicated as potential counterparts of unassociated \fer\ sources or with uncertain nature, we established the blazar-like nature of 8 out the 27 sources analyzed and confirmed 14 classifications.
\end{abstract}

\keywords{methods: statistical - galaxies: active - quasars: general - surveys - radiation mechanisms: non-thermal}

\section{Introduction}
\label{sec:intro}
According to the well assessed unification scenario of the active galactic nuclei \citep[AGN; e.g.,][]{antonucci93,urry95}
blazars are radio loud sources, featuring compact radio cores combined with a ``flat'' radio spectra
that extends from frequencies below $\sim$1GHz \citep[e.g.,][]{ugs3,74mhz,ugs6} up to the sub-millimeter band \citep[e.g.,][]{giommi12}.
They are characterized by a variable, non-thermal, continuum and exhibit a typical double bumped spectral energy distribution (SED),
and represent the largest known population of $\gamma$-ray sources \citep[e.g.,][]{abdo10a,ackermann11} 
proving the most relevant contribution to the extragalactic $\gamma$-ray background \citep[e.g.,][]{mukherjee97,abdo10b}.

Blazars are generally classified on the basis of their optical spectra and divided in two main classes:
i) BL Lac objects, labeled as BZBs according to the nomenclature of the \bzcat\footnote{http://www.asdc.asi.it/bzcat/} \citep{massaro09,massaro11}
when presenting featureless optical spectra and ii) flat spectrum radio quasars (hereinafter BZQs)
having a typical quasar-like optical appearance but also featuring high and variable optical polarization.
In particular, blazars are classified as BZB if the rest-frame equivalent width of their optical features is lower than 5 \AA\
\citep{stickel91,stoke91,laurent99,landoni13}.

As recently discovered using the \wse\ all-sky survey \citep{wright10},
blazars show by peculiar infrared (IR) colors \citep{paper1}
mostly due to their non-thermal continuum that allowed to distinguish them 
from other classes of active galaxies \citep[e.g.,][]{paper2,paper3}.
This IR property was also interpreted as due to the lack of observational 
signatures form a dusty torus in the case of BZBs \citep[e.g.,][]{plotkin12}.

The variable, non-thermal emission of both BZBs and BZQs, extending from radio up to TeV energies,
is interpreted as arising from high-energy particles accelerated in a relativistic jet oriented along to the line of sight, 
whereas relativistic effects amplifies both their luminosity and the amplitude of their variability \citep{blandford78,giommi13}.

Recently, we searched for blazar-like objects as potential counterparts of the 
unidentified $\gamma$-ray sources (UGSs) observed with \fer\ \citep{abdo10a,nolan12}
with several methods based on the IR colors alone \citep{paper4,ugs1} 
or combined with other multifrequency observations, as
radio \citep{ugs5} or X-ray properties \citep{ugs4}.
We also explored the use of low radio frequency observations (i.e., below $\sim$1 GHz)
as an alternative possibility to find blazar-like counterparts \citep[e.g.,][]{ugs3,ugs6} for the UGSs listed in the
second \fer-Large Area Telescope (LAT) catalog \citep[2FGL,][]{nolan12}
in addition to other multifrequency analysis \citep[e.g.,][]{mirabal09,ackermann12,cowperthwaite13,masetti13}. 
All these investigations provided several lists of gamma-ray blazar candidates 
that has to be confirmed and classified via optical spectroscopy. 

Here we investigate the optical spectra of two blazar samples
that lie in the footprint of the spectroscopic catalog of 
Sloan Digital Sky Survey data release 9 \citep[SDSS DR9,][]{ahn12}.
The first sample includes all the \bzcat\ sources that have an uncertain classification, uncertain redshift estimates or have been
classified as BL Lac candidates due to the lack of an optical spectrum in literature \citep{massaro11}.
The second sample lists all the $\gamma$-ray blazar candidates, that were identified as potential counterparts of UGSs 
in our previous analyses, and for which SDSS spectra are now available.
This study is complementary to on going spectroscopic campaigns planned to investigate blazar optical properties 
\citep[e.g.,][]{sbarufatti05,sbarufatti09,plotkin10,paggi14}.

The paper is organized as follows.
In Section~\ref{sec:radius} we present the statistical approach adopted to
determine the optical counterparts of the \bzcat\ sources in the SDSS DR9 catalog
while in Section~\ref{sec:sample} we describe the samples considered in our analysis.
In Section~\ref{sec:results} the results of the spectroscopic analysis are illustrated while,
finally, Section~\ref{sec:conclusions} is devoted to our summary and conclusions.
For our numerical results, we use cgs units unless stated otherwise.

\section{Spatial associations}
\label{sec:radius}
The \bzcat\ was mainly compiled on the basis of radio, optical and X-ray surveys 
and the blazar coordinates reported are not uniform. 
The positional accuracy is generally less than $<$1\arcsec\ but it could reach $\sim$ 5\arcsec,
corresponding to the typical uncertainty on the radio positions of the NVSS,
for those sources with radio flux densities close to the survey limit \citep{condon98}.
Since the positional uncertainties for each source are not reported in the \bzcat\ 
to identify the SDSS optical counterparts of the \bzcat\ blazars we adopted the following a statistical approach.

First we computed the total number of \bzcat\ blazars that lie within the footprint of the SDSS, corresponding to 1820 sources.
For each blazar, we counted the total number of optical counterparts in the SDSS $N(R)$ present
within circular regions of variable radius $R$ in the range between 0\arcsec\ and 10\arcsec.
To be conservative in our analysis, we only included in the $N(R)$ calculation SDSS sources having the flags: 
CLASS\_OBJECT (i.e., mode) and CODE\_MISC (i.e., clean) 
both equal to 1\footnote{http://cas.sdss.org/dr7/sp/help/browser/browser.asp?n=PhotoObj\&t=U}.

We then created 100 mock realizations of the \bzcat\ by shifting each blazar position
in a random direction of the sky by a fixed length of 30\arcsec.
The shift used to create the mock \bzcat\ catalogs were chosen not too distant from the original \bzcat\ location and within the SDSS footprint.
This guarantees to obtain fake catalogs with a sky distribution similar to the original \bzcat\
and to crossmatch each fake catalog and the SDSS taking into account the local density distribution of the optical sources.
The total number of blazars in each mock realization is also preserved being equal to that of the \bzcat\ sources that lie in the SDSS footprint.
For each mock realization of the \bzcat\, we counted the number of associations with the SDSS occurring 
at angular separations $R$ smaller than 10\arcsec. Then we computed the mean number $\lambda(R)$ of these 
fake associations, averaged over the 100 mock \bzcat\ catalogs, verifying that $\lambda(R)$ has a Poissonian distribution.
Increasing the radius by $\Delta\,R=$0\arcsec.2, we also calculated the difference $\Delta\,\lambda(R)$ as:
\begin{equation}
\Delta\,\lambda(R) = \lambda(R+\Delta\,{R}) - \lambda(R)\,,
\end{equation}

\begin{figure}[] 
          \includegraphics[height=9.5cm,width=6.6cm,angle=-90]{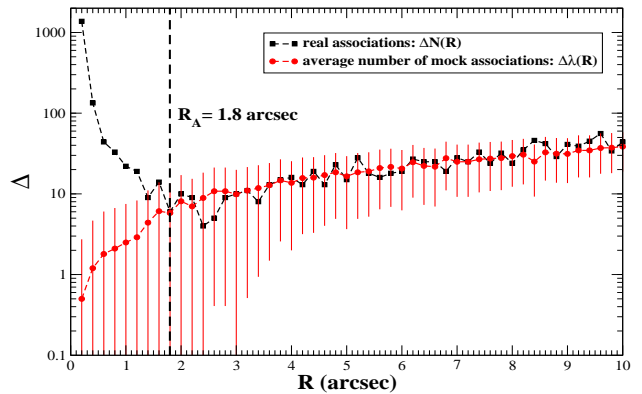}
           \caption{The values of $\Delta\,\lambda(R)$ (red circles) and $\Delta\,N(R)$ (black squares) 
                        as function of the angular separation $R$.
                        We restricted the x axis to $R$ values below 10\arcsec.
                        Our choice of $R_A$ is marked by the vertical dashed line. It occurs at the first 
                        $R$ value for which $\Delta\,\lambda(R)\simeq\Delta\,N(R)$.
                        }
          \label{fig:delta}
          \end{figure}

In Figure~\ref{fig:delta} we show the comparison between $\Delta\,N(R)$ and $\Delta\,\lambda(R)$.
For radii larger than $R_A=$1\arcsec.8 the $\Delta\,\lambda(R)$ curve superimposes that of $\Delta\,N(R)$
indicating that \bzcat-SDSS cross-matches could occur by chance at angular separations larger than $R_A$.
Thus we choose 1\arcsec.8 as to the maximum angular separation between the \bzcat\ and the SDSS position
at which we consider the optical source a reliable counterpart of the blazar in the \bzcat.

Finally the chance probability of spurious associations $p(R_A)$ 
was calculated as the ratio between the number of real associations $N(R_A)$
and the average of those found in the mock realizations of the \bzcat\ $\lambda(R_A)$, corresponding to a value of $\sim$1\%
\citep[see also][for additional details on $p(R_A)$]{maselli10,paper1,ugs1}.

\section{Sample selection}
\label{sec:sample}
We adopted the value of $R_A$ of 1\arcsec.8 to search for the optical counterparts of the blazar-like sources in our two samples 
within the available spectra of the SDSS DR9 to confirm the source nature and whenever possible to estimate the redshift.
Then we analyzed two samples of sources that lie in the footprint of the SDSS dr9 and with optical spectra available 
defined in the following.
 
\begin{enumerate} 
\item The total number of \bzcat\ sources having a optical counterpart in the SDSS DR9 spectroscopic 
catalog within $R_A$ is 219. Among them there are 50 blazars with an uncertain redshift estimate 
and additional 34 sources classified as BL Lac candidates for which optical spectra were not available in literature
while \bzcat\ v4.1 was prepared.
These 84 \bzcat\ sources, with a unique correspondence in the SDSS DR9 spectroscopic catalog within 1\arcsec.8, 
constitute our first sample investigated.

\item The second sample lists 15 blazar-like sources with uncertain classification and/or uncertain $z$ estimates
included in the 2FGL \citep{nolan12} and in the Second \fer-LAT AGN Catalog \citep[2LAC][]{ackermann11}
plus additional 12 sources identified as potential counterparts of UGSs according to their 
peculiar IR colors in our recent analyses \citep[e.g.,][]{paper3,paper4,ugs2}.
All these 27 sources also have a unique correspondence in the SDSS DR9 spectroscopic catalog within 1\arcsec.8.
\end{enumerate}

\section{Results}
\label{sec:results}
We visually inspected all the optical spectra available for the sources in our samples 
to avoid misleading classifications due to artifacts of the SDSS automatic analysis,
and if necessary, for example to confirm a redshift estimate, we downloaded and analyzed the raw data. 

We remark that for the BZB classification we adopted the criterion described in Laurent-Muehleisen et al. (1999),
measuring the rest-frame equivalent widths of the emission and/or absorption lines whenever they are detectable above the continuum
\citep[see also the recent analyses performed by][]{sbarufatti06,landoni13}.
In addition, we also adopted the criterion developed by Massaro et al. (2013c) to classify BL Lac object based on
the SDSS (u-r) color, that supersedes the one based on the Ca H\&K break contrast originally introduced by Stoke et al. (1991).
Thus, for each source, we computed the absorption corrected (u-r) color equal to $(u-r)_{obs}-0.81*A_r$,
where $A_r$ is the Galactic extinction in the R band, and we considered BL Lac objects 
only those sources with $(u-r)<1.4$ (see also Maselli et al. 2013).
We assign a BZB classification only to sources having both a ``featureless'' spectra and the $(u-r)$ color lower than 1.4.

\subsection{\bzcat\ blazars}
\label{sec:blazars}
In the first sample of 84 \bzcat\ blazars with SDSS DR9 spectra, we found that there are 3 BZQs, all associated to \fer\ sources 
in the 2FGL and in the 2LAC \citep{nolan12,ackermann11}, with an uncertain $z$ estimate.
Our analysis of their optical SDSS spectra allowed us to confirm both their nature and their redshifts.
In addition to these BZQs, there are 47 sources out of the 84 listed in the first sample 
classified as BZB according to the \bzcat\ but with an uncertain redshift estimate.
We found that 9 of them have good SDSS spectra from which we obtained a $z$ measurement.
Unfortunately none of these 9 is detected in the $\gamma$-rays.

The remaining 34 sources out of 84 objects in the first sample are indeed classified as BL Lac candidates, 
5 of them being associated to \fer\ sources in the 2LAC.
We confirmed the BL Lac nature for 20 of them, including all the \fer\ sources and providing a new 
redshift estimate for 6. The remaining 14 sources were classified as: normal galaxies (8),
type 2 Seyfert galaxies (5), according to the criteria described in Winkler (1992) 
plus 1 source having still an uncertain nature, mostly resembling a type 2 AGN.

\begin{figure*}[b] 
          \includegraphics[height=9.5cm,width=6.6cm,angle=-90]{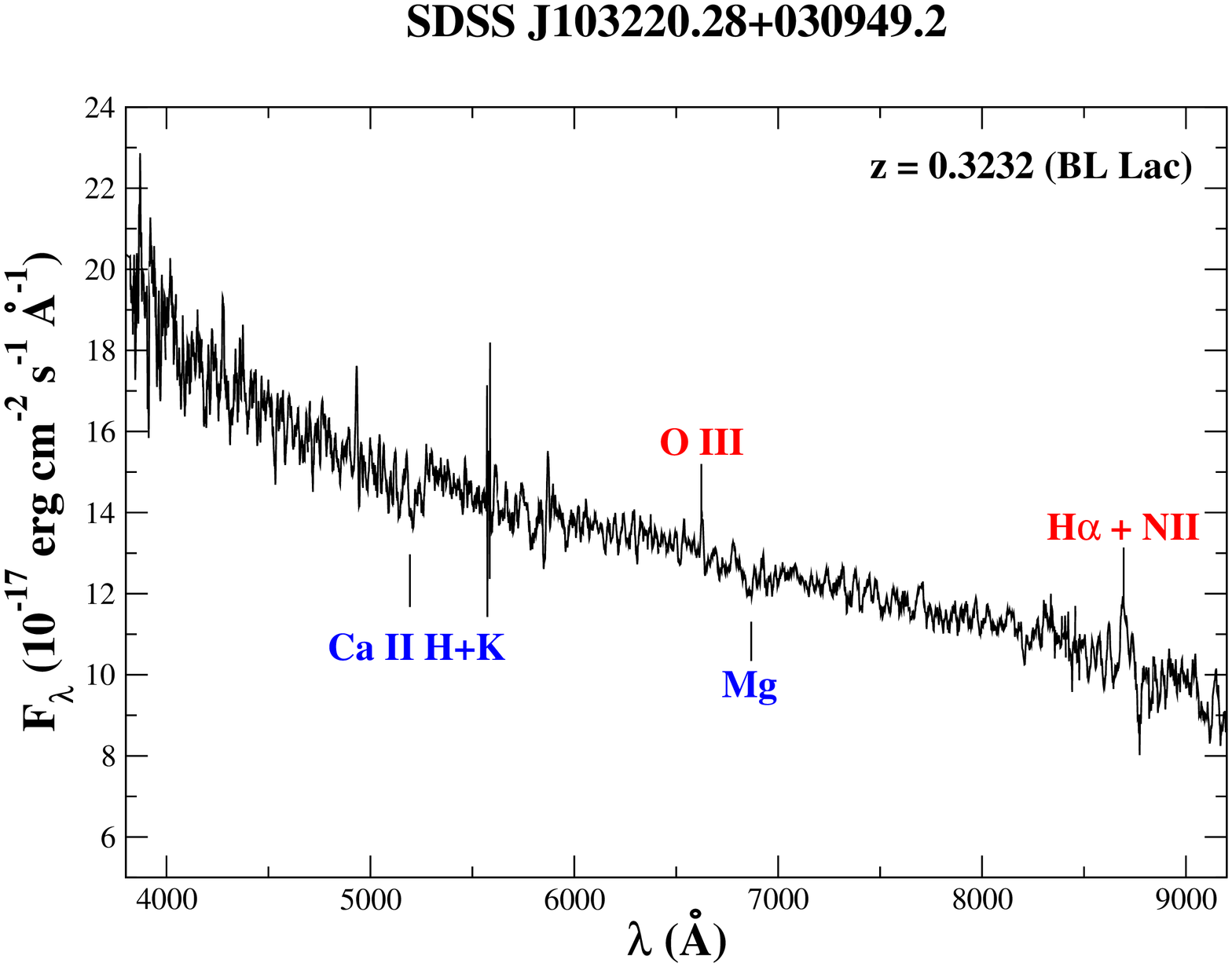}
          \includegraphics[height=9.5cm,width=6.6cm,angle=-90]{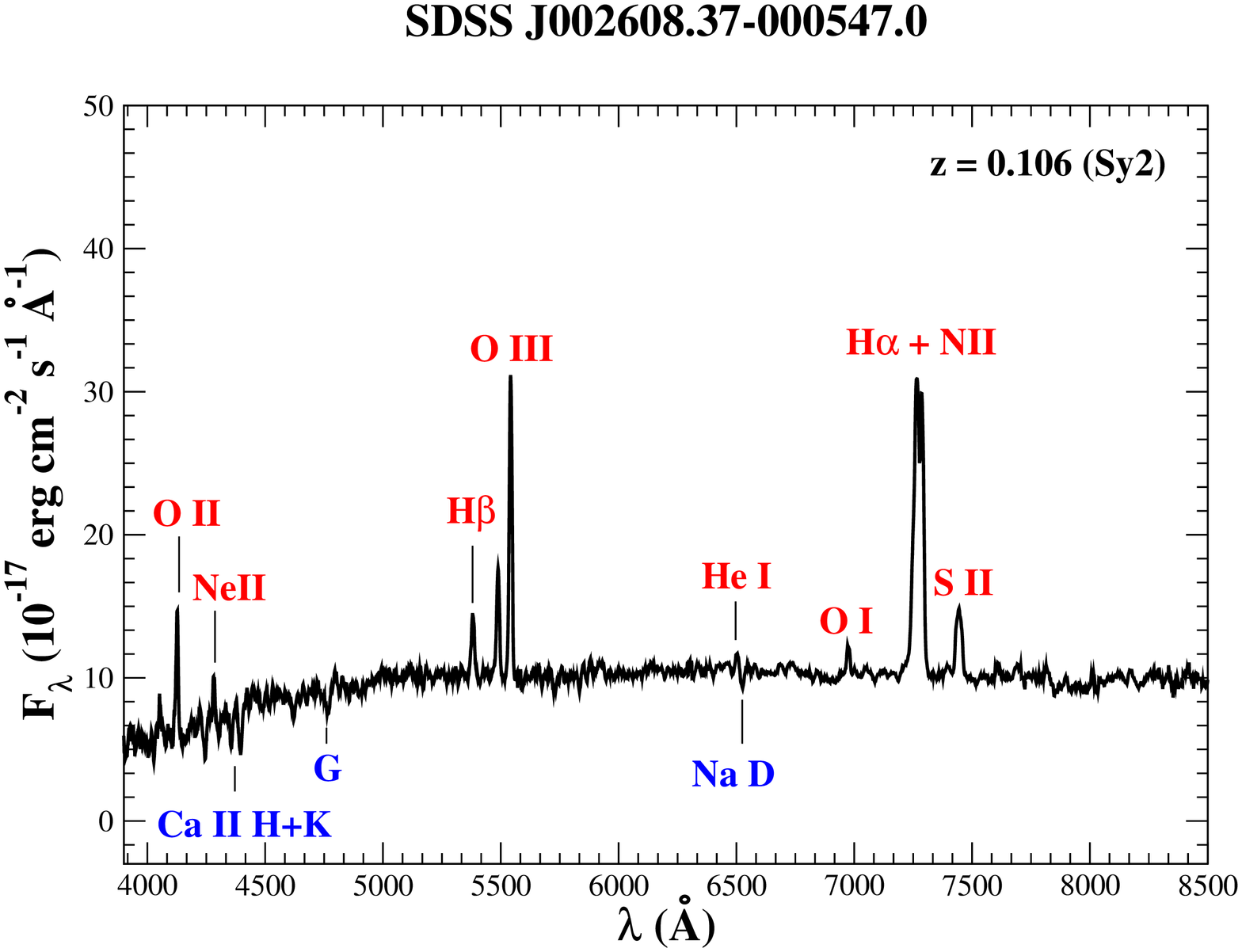}
          \includegraphics[height=9.5cm,width=6.6cm,angle=-90]{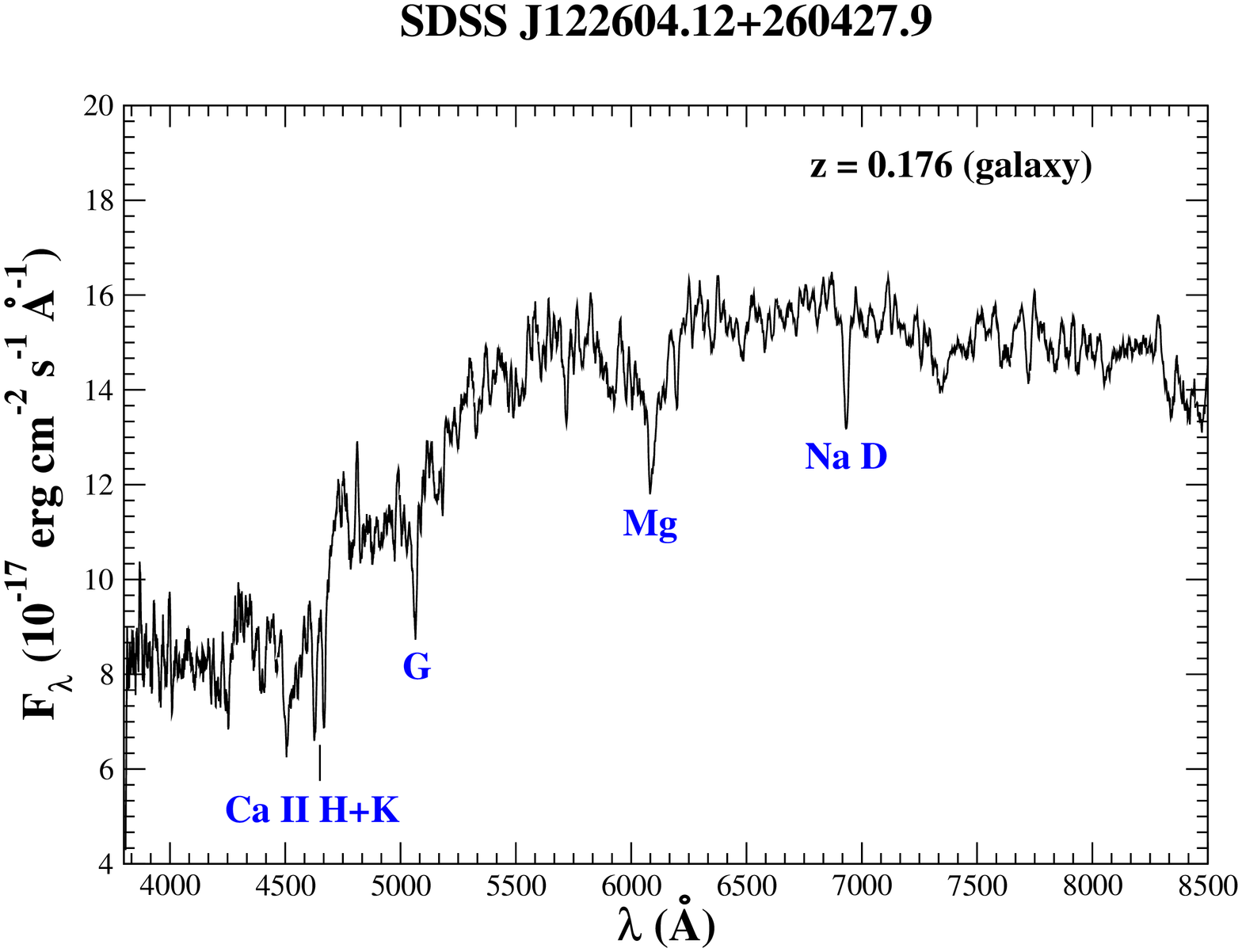}
          \includegraphics[height=9.5cm,width=6.6cm,angle=-90]{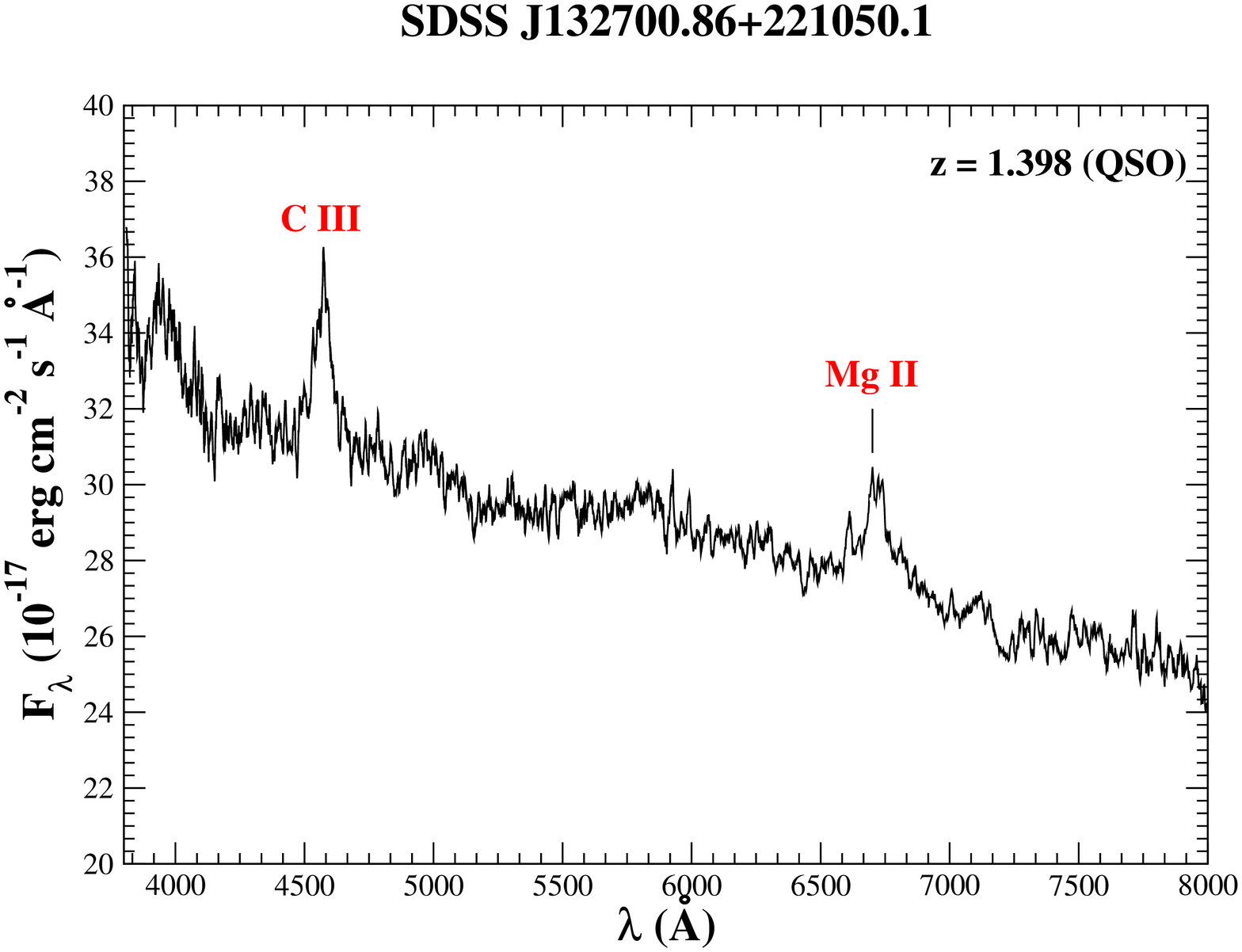}
           \caption{Upper left panel)  The BL Lac object SDSSJ103220.28+030949.2 at redshift 0.3232.
                           Upper right panel) The Seyfert 2 galaxy SDSSJ002608.37-000547.0 previously misclassified as a BL Lac candidate
                           Lower left panel) The normal galaxy SDSSJ122604.12+260427.9 spectrum at redshift z=0.176. 
                           Lower right panel) The z=1.398 quasar SDSSJ132700.86+221050.1.
                           The man in spectral emission and/or absorption features are marked in each figure.
                           }
          \label{fig:spectra}
          \end{figure*}

All our results are summarized in Table~\ref{tab:bzcat}, where we report the \bzcat\ and the SDSS names,
together with the results of our analysis (i.e., classification and redshift estimates when possible) and their (u-r) colors.
Blazars that are associated to \fer\ sources are also indicated.
Uncertain values of redshifts are indicated with a question mark (?); 
they are due to the poor signal to noise of few SDSS archival spectra 
or to the presence of only a single emission/absorption feature.
Then, in Figure~\ref{fig:spectra} we show the optical spectrum of one of the BL Lac classified 
from our analysis together with two cases of wrong classifications and a quasar.

\subsection{$\gamma$-ray blazar candidates}
\label{sec:candidates}
The second sample of $\gamma$-ray blazar candidates selected according to our IR based procedures 
and having SDSS spectra available lists: 27 sources.
Fifteen blazar-like sources were already present in the 2LAC but with uncertain classification or 
uncertain redshift estimates. Among them we found 7 having quasar-like optical spectra
and being classified as BZQs, 2 also with new $z$ estimates, 7 BZBs including 2 sources with measured redshifts  
and 1 misclassified object: SDSS J122011.88+020342.2, associated with 2FGLJ1219.7+0201 
that appears to be a Seyfert galaxy rather than BZQ.

For 2FGLJ1023.6+3947, associated to the SDSS J102333.50+395312.7 source,
the we obtained a redshift of 1.3328 instead of 1.254 reported in the 2LAC 
and for the BL Lac object SDSS J110021.05+401928.0 counterpart of 2FGLJ1100.9+4014,
we were not able to find any optical feature to confirm the 2LAC redshift of 0.225.
Among these 15 sources there is SDSS J222329.57+010226.6,
associated to the AGN of uncertain type 2FGLJ2223.4+0104 and
selected in Cowperthwaite et al. (2013) as a $\gamma$-ray blazar candidate 
that we confirmed as a BL Lac at unknown redshift.

The remaining 12 sources were all selected as $\gamma$-ray blazar candidates
in our previous analyses of their IR colors \citep{paper4,ugs2,ugs3,ugs4}.
We found that 3 sources, all with new $z$ estimates, out of 12 have a quasar like spectrum, 
similar to those of the BZQs, plus one uncertain due to noisy SDSS spectrum
(i.e., SDSS J015852.77+010132.8).
Then there are 4 confirmed BL Lac objects while the remaining 4 sources are indeed 
contaminants of the association methods (1 star and 4 Seyfert galaxies).

All these results are reported in Table~\ref{tab:unknown} in the same order as discussed above,
where we also indicate the previous classification of each source.

\section{Summary and conclusions}
\label{sec:conclusions}
We performed an analysis of the archival optical spectra present in the SDSS DR9 \citep{ahn12}
for two selected samples of blazars and $\gamma$-ray blazar candidates 
to confirm their nature and whenever possible to estimate their redshifts.

First, we adopted a statical approach to find the 
the SDSS optical cross-matches of the sources in our sample.
Then, we analyzed a first sample of 84 blazars listed in the \bzcat\ as BL Lac candidates 
or as BL Lac objects and flat spectrum radio quasars
with uncertain redshift estimates and a second sample of 27 $\gamma$-ray blazar candidates 
selected according to their peculiar IR colors or with uncertain classification 
\citep[e.g.,][and references therein]{ugs1,ugs2,ugs5}.

On the basis of the SDSS spectra, we confirmed the redshifts for 3 flat spectrum radio quasars 
(all $\gamma$-ray sources detected by \fer) and measured the $z$ for 9 additional BL Lacs investigated.
Then, we have been able to classify 34 BL Lac candidates listed in the \bzcat, 
20 of them appearing as BL Lac objects, providing new $z$ estimates for 6 BL Lac objects.
These spectroscopic information, even if available for a small fraction of the whole \bzcat\ catalog
will be essential to refine its future releases as well as those of the \fer\ catalogs.
 
For the second sample listing 27 $\gamma$-ray blazar candidates
we found a total of 11 BZBs (2 with new $z$ measurements) and 11 BZQs having new redshift estimates for 2 of them.
The remaining 5 sources did not appear to have the typical blazar-like optical spectrum. 
All our results are summarized in Table~\ref{tab:summary}.

\acknowledgements
We thank our anonymous referee for many helpful comments which greatly improved this manuscript.
We are grateful to our colleague E. Palazzi that helped us in the data reduction of some SDSS spectra.
The work is supported by the NASA grants NNX12AO97G and NNX13AP20G.
Part of this work is based on archival data, software or on-line services provided by the ASI Science Data Center.
This research has made use of data obtained from the high-energy Astrophysics Science Archive
Research Center (HEASARC) provided by NASA's Goddard
Space Flight Center; the SIMBAD database operated at CDS,
Strasbourg, France; the NASA/IPAC Extragalactic Database
(NED) operated by the Jet Propulsion Laboratory, California
Institute of Technology, under contract with the National Aeronautics and Space Administration.
This publication makes use of data products from the Wide-field Infrared Survey Explorer, 
which is a joint project of the University of California, Los Angeles, and 
the Jet Propulsion Laboratory/California Institute of Technology, 
funded by the National Aeronautics and Space Administration.
TOPCAT\footnote{\underline{http://www.star.bris.ac.uk/$\sim$mbt/topcat/}} 
\citep{taylor05} for the preparation and manipulation of the tabular data and the images.
Funding for the SDSS and SDSS-II has been provided by the Alfred P. Sloan Foundation, 
the Participating Institutions, the National Science Foundation, the U.S. Department of Energy, 
the National Aeronautics and Space Administration, the Japanese Monbukagakusho, 
the Max Planck Society, and the Higher Education Funding Council for England. The SDSS Web Site is http://www.sdss.org/.
The SDSS is managed by the Astrophysical Research Consortium for the Participating Institutions. The Participating Institutions are the American Museum of 
Natural History, Astrophysical Institute Potsdam, University of Basel, University of Cambridge, Case Western Reserve University, University of Chicago, Drexel 
University, Fermilab, the Institute for Advanced Study, the Japan Participation Group, Johns Hopkins University, the Joint Institute for Nuclear Astrophysics, the 
Kavli Institute for Particle Astrophysics and Cosmology, the Korean Scientist Group, the Chinese Academy of Sciences (LAMOST), Los Alamos National 
Laboratory, the Max-Planck-Institute for Astronomy (MPIA), the Max-Planck-Institute for Astrophysics (MPA), New Mexico State University, Ohio State 
University, University of Pittsburgh, University of Portsmouth, Princeton University, the United States Naval Observatory, and the University of Washington.

{}

\begin{table}
\caption{Association of \bzcat\ sources.}
\tiny
\begin{tabular}{|llllccc|}
\hline
  BZCAT  &  SDSS          & BZCAT  & SDSS    & BZCAT    & SDSS    & u-r \\
  name   &  counterpart   & class  & class   & redshift & redshift  & \\
\hline
\noalign{\smallskip}
  BZQJ0310+3814$^*$ & J031049.87+381453.8 & FSRQ & QSO & 0.816? & 0.816 & 1.02 \\
  BZQJ0830+2410$^*$ & J083052.08+241059.8 & FSRQ & QSO & 0.939? & 0.939 & 0.42 \\
  BZQJ1327+2210$^*$ & J132700.86+221050.1 & FSRQ & QSO & 1.4?   & 1.398 & 0.44 \\  
\noalign{\smallskip}
\hline
  BZBJ0001-0011     & J000121.46-001140.3 & BL Lac & BL Lac & 0.462?  & 0.462  & 0.67 \\
  BZBJ0100-0055     & J010058.19-005547.7 & BL Lac & BL Lac & 0.67>?  & ?      & 0.98 \\
  BZBJ0127-0821     & J012716.31-082128.8 & BL Lac & BL Lac & 0.36?   & 0.75?  & 1.06 \\
  BZBJ0141-0928$^*$ & J014125.83-092843.7 & BL Lac & BL Lac & 0.73?   & 0.03?  & 0.96 \\
  BZBJ0731+2804     & J073152.73+280432.9 & BL Lac & BL Lac & 0.25?   & 0.248  & 1.2  \\
  BZBJ0755+3726     & J075523.11+372618.7 & BL Lac & BL Lac & 0.606?  & ?      & 0.9  \\
  BZBJ0801+1336     & J080115.01+133642.2 & BL Lac & BL Lac & 1.042?  & ?      & 0.77 \\
  BZBJ0818+4222$^*$ & J081815.99+422245.4 & BL Lac & BL Lac & 0.53??  & ?      & 0.98 \\ 
  BZBJ0823+2223     & J082324.75+222303.2 & BL Lac & BL Lac & 0.951?  & ?      & 0.87 \\
  BZBJ0840+3440     & J084013.03+344026.9 & BL Lac & BL Lac & 0.451?  & ?      & 1.07 \\
  BZBJ0856+2057$^*$ & J085639.74+205743.3 & BL Lac & BL Lac & 0.18?   & 0.2?   & 0.83 \\
  BZBJ0926+5411     & J092638.87+541126.5 & BL Lac & BL Lac & 0.841?  & 0.8?   & 0.81 \\
  BZBJ0940+2603     & J094014.72+260330.0 & BL Lac & BL Lac & 0.498?? & ?      & 1.11 \\ 
  BZBJ0951+0102     & J095127.81+010210.2 & BL Lac & BL Lac & 0.502?  & ?      & 0.75 \\
  BZBJ1006+3454     & J100656.46+345445.1 & BL Lac & BL Lac & 0.612?  & ?      & 0.49 \\
  BZBJ1012+0630$^*$ & J101213.34+063057.1 & BL Lac & BL Lac & 0.518?  & ?      & 0.69 \\
  BZBJ1031+5053$^*$ & J103118.51+505335.8 & BL Lac & BL Lac & 0.361?? & ?      & 0.45 \\ 
  BZBJ1032+0309     & J103220.28+030949.2 & BL Lac & BL Lac & 0.323?  & 0.3232 & 0.64 \\
  BZBJ1100+4019$^*$ & J110021.05+401928.0 & BL Lac & BL Lac & 0.225?  & ?      &-0.02 \\
  BZBJ1107+5010     & J110704.78+501037.9 & BL Lac & BL Lac & 0.706?  & 0.7062 & 0.99 \\
  BZBJ1117+2548     & J111740.39+254846.5 & BL Lac & BL Lac & 0.36?   & ?      & 0.63 \\
  BZBJ1120+4212$^*$ & J112048.06+421212.4 & BL Lac & BL Lac & 0.124?? & ?      & 0.34 \\ 
  BZBJ1132+0034$^*$ & J113245.62+003427.7 & BL Lac & BL Lac & 1.223?? & ?      & 0.89 \\ 
  BZBJ1136+1601     & J113617.53+160152.2 & BL Lac & BL Lac & 0.574?  & 0.5734 & 0.8  \\
  BZBJ1138+4113     & J113812.15+411352.8 & BL Lac & BL Lac & 0.574?  & 0.5740?& 1.1  \\
  BZBJ1211+2242     & J121158.63+224232.9 & BL Lac & BL Lac & 0.455?  & 0.4527 & 0.53 \\
  BZBJ1219+0446     & J121944.97+044622.4 & BL Lac & BL Lac & 0.489?  & ?      & 0.72 \\
  BZBJ1231+2847$^*$ & J123143.57+284749.7 & BL Lac & BL Lac & 0.236?  & ?      & 0.68 \\
  BZBJ1237+3020     & J123705.61+302005.1 & BL Lac & BL Lac & ?       & ?      & 0.36 \\ 
  BZBJ1238+4431     & J123826.01+443137.1 & BL Lac & BL Lac & 0.312?  & 0.3121 & 1.34 \\
  BZBJ1239+4132     & J123922.73+413251.4 & BL Lac & BL Lac & 0.16??  & ?      & 0.53 \\ 
  BZBJ1247+4423     & J124700.72+442318.8 & BL Lac & BL Lac & 0.6?    & 0.73?  & 0.51 \\
  BZBJ1255+3848     & J125555.40+384811.3 & BL Lac & BL Lac & 0.559?  & ?      & 0.81 \\
  BZBJ1328+1145     & J132833.56+114520.5 & BL Lac & BL Lac & 0.49?   & ?      &-0.06 \\
  BZBJ1401+3611     & J140138.72+361121.9 & BL Lac & BL Lac & 0.507?  & 0.5064?& 1.21 \\
  BZBJ1404+0402$^*$ & J140450.90+040202.1 & BL Lac & BL Lac & 0.344?? & ?      & 0.45 \\ 
  BZBJ1423+1412     & J142330.67+141247.9 & BL Lac & BL Lac & 0.769?? & 0.7687?& 0.6  \\ 
  BZBJ1436+4129     & J143627.16+412932.3 & BL Lac & BL Lac & 0.404?  & ?      & 2.74 \\
  BZBJ1436+5639$^*$ & J143657.72+563924.8 & BL Lac & BL Lac & 0.15?   & ?      & 0.47 \\
  BZBJ1443+2515     & J144334.40+251558.2 & BL Lac & BL Lac & 0.529?? & 0.5295 & 0.8  \\ 
  BZBJ1456+5048     & J145603.64+504825.9 & BL Lac & BL Lac & 0.479?  & ?      & 0.52 \\
  BZBJ1506+0814$^*$ & J150644.47+081400.6 & BL Lac & BL Lac & 0.376?  & ?      & 0.52 \\
  BZBJ1553+0601     & J155331.06+060143.8 & BL Lac & BL Lac & 0.619?  & 0.6189 & 0.98 \\
  BZBJ1603+1105     & J160341.93+110548.7 & BL Lac & BL Lac & 0.143?? & ?      & 0.94 \\ 
  BZBJ1623+2841     & J162332.25+284128.7 & BL Lac & BL Lac & 0.377?  & ?      & 0.46 \\
  BZBJ1652+3632     & J165248.44+363212.5 & BL Lac & BL Lac & 0.648?  & 0.6470?& 0.96 \\
  BZBJ1701+3954     & J170124.63+395437.0 & BL Lac & BL Lac & 0.507?  & ?      & 0.27 \\
\noalign{\smallskip}
\hline
\end{tabular}\\
Col. (1) \bzcat\ name. \\
Col. (2) SDSS name of the optical counterpart. \\
Col. (3) \bzcat\ classification. \\
Col. (4) SDSS spectroscopic classification based on our analysis. \\
Col. (5) Redshift estimate reported in the \bzcat. Question mark indicates uncertain estimates.\\
Col. (6) Redshift estimate derived from our analysis.Question mark indicates uncertain estimates.\\
Col. (7) SDSS u-r color.\\
\label{tab:bzcat}
\end{table}

\begin{table}
\caption{Association of \bzcat\ sources.}
\tiny
\begin{tabular}{|llllcccc|}
\hline
  BZCAT  &  SDSS          & BZCAT  & SDSS    & BZCAT    & SDSS    & u-r & Classification\\
  name   &  counterpart   & class  & class   & redshift & redshift  & & flag\\
\hline
\noalign{\smallskip}
  BZBJ0026-0005     & J002608.37-000547.0 & BL Lac Can. & Sy2    & 0.107  & 0.106  & 2.16 & no  \\
  BZBJ0109+1816$^*$ & J010908.17+181607.5 & BL Lac Can. & BL Lac & 0.145  & ?      & 0.7  & yes \\
  BZBJ0253-0124     & J025315.60-012405.3 & BL Lac Can. & BL Lac & ?      & ?      & 0.67 & yes \\
  BZBJ0754+4823$^*$ & J075445.66+482350.7 & BL Lac Can. & BL Lac & ?      & ?      & 0.93 & yes \\
  BZBJ0814+0856     & J081421.66+085706.1 & BL Lac Can. & galaxy & 0.23?? & 0.24   & 7.33 & no  \\ 
  BZBJ0829+1754     & J082904.82+175415.8 & BL Lac Can. & Sy2    & 0.089  & 0.0895 & 2.12 & no  \\
  BZBJ0831+5400     & J083100.36+540023.2 & BL Lac Can. & Sy2    & ?      & 0.0617 & 3.06 & no  \\
  BZBJ0839+4015     & J083903.08+401545.6 & BL Lac Can. & galaxy & 0.194  & 0.1941 & 2.43 & no  \\
  BZBJ0905+4705     & J090536.44+470546.3 & BL Lac Can. & type2  & 0.174  & 0.1736 & 2.33 & no  \\
  BZBJ0912+4235     & J091227.22+423545.1 & BL Lac Can. & galaxy & 0.266  & 0.2662 & 4.0  & no  \\
  BZBJ0933+0003     & J093310.57+000323.5 & BL Lac Can. & BL Lac & ?      & ?      & 0.45 & yes \\
  BZBJ0944+5557     & J094441.47+555752.9 & BL Lac Can. & BL Lac & ?      & ?      & 0.93 & yes \\
  BZBJ1007+5023     & J100710.44+502356.4 & BL Lac Can. & Sy2    & 0.133  & 0.1326 & 1.87 & no  \\
  BZBJ1057+2303     & J105723.09+230318.7 & BL Lac Can. & BL Lac & 0.379  & 0.3782 & 1.07 & yes \\
  BZBJ1058+2817     & J105829.89+281746.3 & BL Lac Can. & BL Lac & ?      & 0.4793?& 0.89 & yes \\
  BZBJ1100+4210     & J110020.99+421053.1 & BL Lac Can. & galaxy & 0.323  & 0.3229 & 1.75 & no  \\
  BZBJ1110+3539     & J111056.83+353907.2 & BL Lac Can. & BL Lac & ?      & 0.61?  & 0.72 & yes \\
  BZBJ1111+3452     & J111130.90+345203.2 & BL Lac Can. & BL Lac & 0.212  & ?      & 0.42 & yes \\
  BZBJ1152+2837     & J115210.70+283721.3 & BL Lac Can. & BL Lac & ?      & 0.4412 & 1.17 & yes \\
  BZBJ1153+3823     & J115210.70+283721.3 & BL Lac Can. & Sy2    & ?      & 0.4098 & 0.99 & no  \\
  BZBJ1224+2239     & J122401.03+223939.5 & BL Lac Can. & BL Lac & ?      & 0.4821 & 0.88 & yes \\
  BZBJ1226+2604     & J122604.12+260427.9 & BL Lac Can. & galaxy & 0.176  & 0.1761 & 2.0  & no  \\
  BZBJ1243+3627$^*$ & J124312.73+362743.9 & BL Lac Can. & BL Lac & ?      & ?      & 0.45 & yes \\
  BZBJ1253+3826     & J125300.95+382625.7 & BL Lac Can. & BL Lac & 0.372  & 0.3707 & 1.08 & yes \\
  BZBJ1311+0853     & J131155.76+085340.9 & BL Lac Can. & BL Lac & 0.469  & 0.4694 & 0.85 & yes \\
  BZBJ1314+2348$^*$ & J131443.80+234826.7 & BL Lac Can. & BL Lac & ?      & 0.15?  & 0.72 & yes \\
  BZBJ1341+3716     & J134138.66+371644.8 & BL Lac Can. & galaxy & 0.17   & 0.1745 & 2.93 & no  \\
  BZBJ1404+2701     & J140436.82+270141.0 & BL Lac Can. & galaxy & 0.136  & 0.1383 & 2.67 & no  \\
  BZBJ1410+2820$^*$ & J141029.56+282055.6 & BL Lac Can. & BL Lac & ?      & ?      & 0.58 & yes \\
  BZBJ1426+2415     & J142645.52+241523.0 & BL Lac Can. & BL Lac & 0.055??& 0.36?  & 0.4  & yes \\ 
  BZBJ1437+4717     & J143716.14+471726.3 & BL Lac Can. & BL Lac & ?      & ?      & 0.68 & yes \\
  BZBJ2129+0035     & J212940.67+003527.4 & BL Lac Can. & BL Lac & 0.425  & 0.4264 & -0.1 & yes \\
  BZBJ2227+0037     & J222758.13+003705.4 & BL Lac Can. & BL Lac & ?      & ?      & 0.76 & yes \\
  BZBJ2319-0116     & J231952.83-011626.8 & BL Lac Can. & galaxy & ?      & 0.2835 & 1.72 & no  \\
\noalign{\smallskip}
\hline
\end{tabular}\\
Col. (1) \bzcat\ name. \\
Col. (2) SDSS name of the optical counterpart. \\
Col. (3) \bzcat\ classification. \\
Col. (4) SDSS spectroscopic classification based on our analysis. \\
Col. (6) Redshift estimate derived from our analysis.Question mark indicates uncertain estimates.\\
Col. (7) u-r color .\\
Col. (8) Classification flag: (yes) marks sources that have been classified on the basis of our analysis.\\
\label{tab:bzcat}
\end{table}

\begin{table}
\caption{Association of $\gamma$-ray blazar candidates (00-24 HH).}
\tiny
\begin{tabular}{|lllccc|}
\hline
  Source  &  SDSS          & SDSS    & SDSS     & u-r & Classification\\
   name    &  counterpart  & class   & redshift & & flag \\
\noalign{\smallskip}
\hline
  2FGLJ0323.6-0108 &  J032343.62-011146.1 & BL Lac & ?      & 0.58 & yes  \\ 
  2FGLJ0924.0+2819 &  J092351.52+281525.1 & QSO    & 0.7442 & 0.57 & yes  \\ 
  2FGLJ0950.1+4554 &  J095011.82+455320.0 & BL Lac & 0.3994 & 0.99 & yes  \\ 
  2FGLJ1017.0+3531 &  J101810.97+354239.4 & QSO    & 1.2280 & 0.42 & yes  \\ 
  2FGLJ1023.6+3947 &  J102333.50+395312.7 & QSO    & 1.3328 & 0.23 & yes  \\ 
  2FGLJ1100.9+4014 &  J110021.05+401928.0 & BL Lac & ?      & 0.43 & yes  \\ 
  2FGLJ1219.7+0201 &  J122011.88+020342.2 & Sy1.8  & 0.2402 &-0.02 & no  \\ 
  2FGLJ1222.4+0413 &  J122222.54+041315.7 & QSO    & 0.9642 & 0.44 & yes  \\ 
  2FGLJ1301.6+3331 &  J130129.15+333700.3 & QSO    & 1.0084 & 0.75 & yes  \\ 
  2FGLJ1310.9+0036 &  J131106.47+003510.0 & BL Lac & ?      & 0.62 & yes  \\ 
  2FGLJ1351.4+1115 &  J135120.84+111453.0 & BL Lac & ?      & 0.56 & yes  \\ 
  2FGLJ1332.7+4725 &  J133245.24+472222.6 & QSO    & 0.6687 & 0.74 & yes  \\ 
  2FGLJ1442.0+4352 &  J144207.15+434836.7 & BL Lac & ?      & 0.78 & yes  \\ 
  2FGLJ1522.0+4348 &  J152149.61+433639.2 & QSO    & 2.1677 & 0.35 & yes  \\ 
  2FGLJ2223.4+0104$^+$ &  J222329.57+010226.6 & BL Lac & 0.29?  & 0.49 & yes  \\ 
\noalign{\smallskip}
\hline
\noalign{\smallskip}
  2FGLJ0158.4+0107 &  J015852.77+010132.8 & QSO?   & 1.61?  & 0.95 & yes  \\
  2FGLJ0440.5+2554 &  J043947.48+260140.8 & star   & 0.     &-1.56 & no  \\
  2FGLJ0823.0+4041 &  J082257.55+404149.7 & QSO    & 0.8655 & 0.58 & yes  \\
  1FGLJ0835.4+0936 &  J083543.21+093717.9 & BL Lac & 0.35?  & 0.79 & yes  \\ 
  2FGLJ0844.9+6214 &  J084406.82+621458.4 & Sy1.9  & 0.1208 & 3.13 & no  \\
  2FGLJ1129.5+3758 &  J112903.24+375656.7 & BL Lac & ?      & 1.14 & yes  \\
  2FGLJ1209.6+4121 &  J120922.78+411941.3 & BL Lac & ?      & 0.75 & yes  \\ 
  1FGLJ1422.7+3743 &  J142304.61+373730.5 & BL Lac & ?      & 0.88 & yes  \\ 
  2FGLJ1612.0+1403 &  J161118.10+140328.7 & QSO    & 0.5855 & 0.15 & yes  \\
  2FGLJ1614.8+4703 &  J161541.21+471111.7 & Sy2    & 0.1986 & 1.31 & no  \\
  2FGLJ1627.8+3219 &  J162800.39+322414.0 & QSO    & 0.9051 & 0.69 & yes  \\
  1FGLJ2117.8+0016 &  J211817.39+001316.7 & Sy1.5  & 0.4629 & 0.79 & no  \\ 
\noalign{\smallskip}
\hline
\end{tabular}\\
Col. (1) Source name. \\
Col. (2) SDSS name of the optical counterpart. \\
Col. (3) SDSS spectroscopic classification based on our analysis. \\
Col. (4) Redshift estimate derived from our analysis. Question mark indicates uncertain estimates.\\
($^+$): 2FGLJ2223.4+0104 is the source indicated by Cowperthwaite et al. (2013).\\
Col. (6) Classification flag: (yes) marks sources that have been classified on the basis of our analysis.\\
\label{tab:unknown}
\end{table}

\begin{table}
\caption{Summary.}
\begin{tabular}{|lrrrrr|}
\hline
  Sample  & Tot. &  BZBs   & BZQs    & candidates & new\,$z$ \\
\noalign{\smallskip}
\hline
(1$^{st}$): \bzcat\ & 84 & 47 (47) & 3 (3) & 34 (20) & 18 \\
(2$^{nd}$): $\gamma$-ray blazar candidates & 27 & 7 (7) & 8 (7) & 12 (4) & 4 \\
\noalign{\smallskip}
\hline
\noalign{\smallskip}
\end{tabular}\\
Col. (2) Total umber of sources listed in the sample.\\
Col. (3) Number of sources classified as BZBs; 
those confirmed by our analysis are in parenthesis.\\
Col. (4) Number of sources classified as BZQs; 
those confirmed by our analysis are in parenthesis.\\
Col. (5) Number of blazar candidates; 
those confirmed by our analysis are in parenthesis.\\
Col. (6) Sources with new $z$ estimates.
\label{tab:summary}
\end{table}

\end{document}